# Aphelion Cloud Belt Phase Function Investigations with Mars Color Imager (MARCI)

*Brittney A. Cooper*[a,d], *John E. Moores*[a], *Joseph M. Battalio*[b], *Scott D. Guzewich*[c], *Christina L. Smith*[a], *Rachel C. N. Modestino*[a], *Michael V. Tabascio*[a]

[a] *Centre for Research in Earth and Space Science, York University, 4700 Keele Street Toronto, M3J1P3, ON, Canada*
[b] *Harvard – Smithsonian Center for Astrophysics, 60 Garden Street, Cambridge, Massachusetts, 02138, United States of America*
[c] *NASA Goddard Space Flight Center, 8800 Greenbelt Rd, Greenbelt, MD 20771, USA*
[d] *Now at Gemini North Observatory, 670 N Aohoku Pl, Hilo, HI, 96720, USA*

**Corresponding Author:** Brittney Cooper
Address: Centre for Research in Earth and Space Science, York University, 4700 Keele St, Toronto, M3J1P3, ON, Canada, Email Address: bcooper@gemini.edu

**Abstract:** This paper constrains the scattering phase function of water ice clouds (WICs) found within Mars' Aphelion Cloud Belt (ACB), determined from orbit by processing publicly available raw Mars Color Imager (MARCI) data spanning solar longitudes ($L_S$) 42°-170° during Mars Years (MYs) 28 and 29. MARCI visible wavelength data were calibrated and then pipeline-processed to select the pixels most likely to possess clouds. Mean phase function curves for the MARCI blue filter data were derived, and for all seasons investigated, modeled aggregates, plates, solid and hollow columns, bullet rosettes, and droxtals were all found to be plausible habits. Spheres were found to be the least plausible, but still possible. Additionally, this work probed the opposition surge to examine the slope of the linear relationship between column ice water content and cloud opacity on Mars, and found a significant dependence on particle radius. The half-width-half-maxima (HWHM) of the visible 180° peak of five MARCI images were found to agree better with modeled HWHMs for WICs than with modeled HWHM for dust.

# 1. Introduction

### 1.1 The Aphelion Cloud Belt

Although 10,000 times less abundant in concentration, on Mars than Earth, water vapour is a powerful and dynamic trace gas in the Martian atmosphere (Maltagliati et al., 2011). Approximately ten percent of the total water vapour in the Martian atmosphere contributes to the seasonally-driven formation of water ice clouds (WICs) in equatorial regions, as well as to the formation of WICs driven by atmospheric waves and dynamics near the poles (Maltagliati et al., 2011). The lifetimes of the constituent ice crystals within these Martian WICs allow for large-scale advection on the order of thousands of kilometers (Montmessin et al., 2004).

Mars' orbital eccentricity of 0.0934 (Simon et al., 1994) has a noticeable impact on its seasonal meteorology and climate. As Mars approaches its greatest distance from the Sun at aphelion around a solar longitude ($L_s$) of $L_s = 71°$, the planet's atmosphere cools and a decrease in dust lifting is observed along with a marked increase in cloud formation and density. Figure 8 of Smith (2008) demonstrates the seasonal peak in cloud opacities spanning approximately $L_s$=50°-180° over multiple Martian years (MY). As Mars approaches the northern hemisphere's autumnal equinox at $L_s$ =180°, the atmosphere begins to warm and more vigorous wind-stress dust lifting resumes, while cloud formation declines (Newman et al., 2005). This decrease in cloud formation continues as Mars reaches and surpasses its closest approach of the Sun at perihelion, at $L_s$=250°. This process repeats annually and gives rise to the phenomenon known as the Aphelion Cloud Belt (ACB): the highest annual concentration of Martian WICs that are typically located within -10° and +30° latitude, with increased opacities typically ranging from 0.05-0.5 as observed in the wavelength range: 410-673 nm (Wolff et al., 1999; Clancy et al. 1996).

## 1.2 Cirrus Clouds as an Analog for Martian WICs

Parallels can be drawn between Terrestrial cirrus clouds and Martian WICs, as both are clouds composed of frozen water ice deposited onto cloud condensation nuclei (CCN) for temperatures well below 273 K (Whiteway et al., 2009). They both form in similar conditions, as the Martian troposphere has similar temperatures and pressures to the Terrestrial stratosphere (Petrosyan et al., 2011). On Earth, cirrus clouds have a significant impact on the global radiation budget as they contribute to a warming greenhouse effect by absorbing infrared (IR) radiation emitted from the Earth's surface and re-emitting it into the atmosphere. They also reflect a substantial amount of solar radiation back out to space, which produces a cooling effect (Poetzsch-Heffter et al., 1995). These effects are altitude-dependent, but when the sum of these effects is tallied, optically thin cirrus clouds are found to be the only Terrestrial clouds that have a net warming effect on Earth's radiation budget.

Similar effects from Martian WICs have been investigated and modeled, generally finding net cooling below 15 km and net warming above (Madeleine et al., 2012). Schlimme et al. (2005) demonstrated that in the case of Terrestrial water ice clouds, the most sensitive parameters for modelling the solar broadband radiative transfer (RT) of clouds are (in order of importance): optical thickness, ice crystal shape, ice particle size, and spatial structure. If any of these parameters are not well constrained, the accuracy of the model will be impaired.

## 1.3 Constraining the Geometries of Ice Crystals Within Martian WICs

The optical depth of Martian WICs is the most sensitive parameter for the modelling of solar broadband RT of clouds, and therefore has been monitored regularly from the surface by the Mars Exploration Rovers (Lemmon et al., 2015), Mars Science Laboratory (MSL) (Moores et al., 2015; Kloos et al., 2016; Kloos et al., 2018), and the Phoenix lander (Moores et al., 2011), and from orbit by the Thermal Emission Imaging Spectrometer (THEMIS) aboard Mars Odyssey (Smith, 2009), the Thermal Emission Spectrometer (TES) aboard Mars Global Surveyor (MGS) (Clancy et al., 2003), and the Mars Climate Sounder (MCS) aboard the Mars Reconnaissance Orbiter (MRO) (Guzewich et al., 2017, Kleinböhl et al., 2009). Ice crystal geometries on the other hand (ranked second in importance because they provide important scattering and phase function information), can be inferred from a Martian WIC scattering phase function analysis (Greenler, 1980) in lieu of in-situ observations. Phase function investigations such as Pollack et al. (1979), Clancy and Lee, (1991), Clancy et al., (2003), and Wolff et al. (2009) have utilized a method of fitting RT models to emission phase function (EPF) data taken over a finite range and resolution of emission angles from various spacecraft imagers. Surprisingly, the majority of these investigations produced relatively flat convex curves devoid of the many local maxima present in modeled ice crystal phase functions by Chepfer et al. (2002), Yang and Liou (1996), and Yang et al. (2010), including the theoretically predicted 180° backscatter peak (Curran et al., 1978; Greenler, 1980). The local maxima that occur in the modeled pure ice phase functions are seemingly important as they correspond to optical effects commonly observed in Terrestrial cirrus clouds, such as halos, arcs, parhelia, and the anti-solar point or opposition (180º backscatter) surge.

  The 180° peak has been observed consistently in orbital images from both the Mars Color Imager (MARCI) and the Mars Orbiter Camera (MOC) (Cooper and Moores, 2019; Wang and Ingersoll, 2002), meaning that it should also logically appear as a peak in the scattering phase function. Cooper et al. (2019) observationally constrained the phase function of Martian WICs from the surface with Mars Science Laboratory (MSL) by observing scattering at visible and NIR wavelengths using the Navcams in contrast to the EPF work of Pollack et al. (1979), Clancy and Lee, (1991), Clancy et al., (2003), and Wolff et al. (2009). While some derived phase function curves for Terrestrial clouds have been relatively flat in comparison to these model outputs, Zhou and Yang (2015) were able to show that the theoretical backscatter peak is entirely consistent with observations of randomly oriented hexagonal columns and plates, and thus should not be neglected. Indeed, Cooper et al. (2019) found that five ice crystal geometries, common in Terrestrial cirrus clouds, were plausible constituents of the clouds observed during 35 weeks of the Mars Year (MY) 34 ACB.

As such, the goal of this study is to extend the work of Cooper et al. (2019), to further constrain the dominant geometries of water ice crystals in Martian WICs from publicly available orbital images captured by MARCI aboard MRO during its Primary Science Phase (PSP). The extension of the phase function investigation from the solely ground-based MSL data of Cooper et al. (2019) to MARCI's PSP data, allowed the range of observations of clouds to be extended from a single Aphelion season observed from a single location within a crater in a single waveband, to global coverage over two aphelion seasons. It was important to expand beyond the MSL observations within Gale crater, as the local meteorology and dynamics within the crater produce a microclimate that is not necessarily indicative of other regions on the Martian surface at similar latitudes (Miller et al., 2018). The temporal and spatial variation of the phase function through the analysis period was also examined , as was the potential for dust contamination in our observations of Martian WICs (Section 3). These phase functions were then used to constrain the seasonal dominant ice crystal geometries in the ACB seasons of MYs 28 and 29, in addition to examining the properties of the WIC 180° backscatter peak (Section 4). The phase function at 180° was used to appraise the use of a Terrestrial empirical relationship for deriving ice water content (IWC ) from water ice extinction or opacity from Dickinson et al. (2011), and to compare the visible half-width-half-maxima (HWHM) of five composite MARCI images to HWHM modeled for water ice crystals and dust and the Martian surface.

## 2. Methods

### 2.1 Radiometric Calibration

MARCI was launched aboard MRO which entered its PSP in orbit around Mars in November 2006. The PSP lasted until November 2008, spanning MY 28 ~$L_s$=128° to MY 29 ~$L_s$=165°. This study used publicly available raw MARCI data (dataset ID: MRO-M-MARCI-2-EDR-L0-V1.0; Malin et al., 2001) taken from the Planetary Data System (PDS; Eliason et al., 1996), captured during the aphelion seasons (~$L_s$=42°-170°) within the PSP. The boundaries of the solar longitude range of interest were selected based upon the results of the Kloos et al. (2018) analysis of ACB cloud optical depths as observed from MSL.

During the PSP, MRO was typically in a 3am/3pm Sun-synchronous orbit (Figure 1a) that allowed MARCI to capture 12 to 13 images per sol, in five visible and two ultraviolet wavelength filters (Zurek and Smrekar, 2007). These filters were permanently mounted on top of MARCI's 180° field of view (FOV) charge coupled device (CCD), operating as a "push broom" imager. MARCI captured frames at regularly timed intervals in each orbit (Bell et al., 2009), and the resultant raw data product was a long multi-filtered swath made up of individual frames captured along the orbit (Figure 1b), with each frame containing five single filter "framelets." The visible (VIS) filters consist of blue (437 $\pm$ 32 nm), green (546 $\pm$ 40 nm), orange (604 $\pm$ 31 nm), red (653 $\pm$ 42 nm), and near infrared (NIR, 750 $\pm$ 50 nm) filters, each with an approximate down-track FOV of 2° and cross-track FOV of 180° (Bell et al. 2009). Due to their permanent placement on the CCD, the FOV and viewing geometries of each MARCI filter are offset from one another.

The VIS blue filter data products were processed using a pipeline that was produced to calibrate the data according to the methods of Bell et al. (2009). The initial calibration consisted of decompanding and flat-fielding the images, taking into account the exposure times, the filter wavelengths, and the distance from the Sun to convert the raw values into spectral radiance and reflectance. Each multi-filter image was reduced in resolution through 8x8 pixel summing to reduce processing time, separated into the five VIS filters, and then run through the pipeline. The blue single filter images that were output were then cropped in width and length to exclude areas with high intensity limb scattering, and polar latitudes where ice is often condensed onto the surface. This cropping was necessary so that the pixels with maximum reflectance values in each column of a cropped image could be assumed to contain Martian WICs (and not ice). The atmospheric limb was removed because it often has opacities too high for the single scattering approximation, and thus cannot be used in this determination of the phase function. The selected maximum reflectance pixels were isolated and used as the inputs for the remainder of the analysis. During the ACB season, it was reasonable to assume that the brightest pixels within the blue filter were likely to be clouds, once the limb and polar regions were removed from the images. This approach is comparable to that taken by Wolff et al. (2010) and we have validated our work against their results (Figure 2). Martian WICs have an average single scattering albedo (SSA) of 1.0 (Clancy et al., 2003) compared to Martian dust with a SSA of 0.75 for blue (Wolff et al., 2009). This analysis was originally done for all 5 VIS filters, but there was insufficient contrast between the clouds and surface for the green, orange, red, and NIR filters, which inhibited this method from being consistently effective in those wavelengths. Thus, this work focuses solely on the blue filter data.

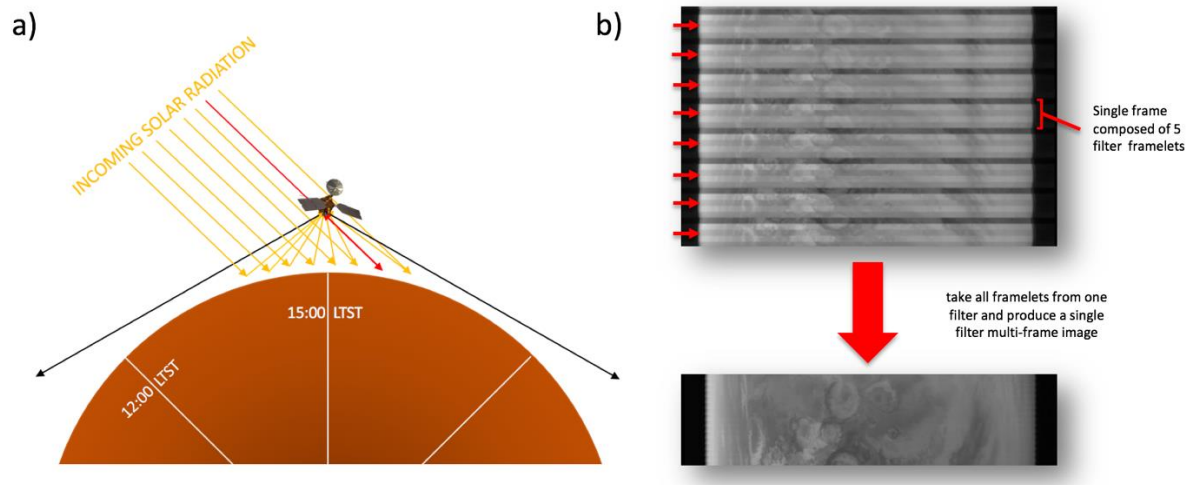

*Figure 1: Panel (a) depicts a rough schematic of the typical viewing geometry of MARCI with MRO in its 3 am/3 pm Sun-synchronous orbit, where it observes the equator at a local true Solar time (LTST) of 15:00. With this geometry, MARCI is able to observe the 180° scattering angle in its filters' FOVs (highlighted in this simple example by the red ray) and capture the 180° backscatter peak around the equatorial region (Cooper and Moores, 2019). Panel (b) displays the format of a raw multi-filter MARCI VIS image downloaded from the PDS, which then has to be separated into its five VIS filters to produce five single filter images, before it can be radiometrically calibrated.*

To validate the calibration, the cross-track reflectance (I/F) obtained from our pipeline for the blue filter of image P10_004770_2885_MU_00N178W, was compared to that of Wolff et al. (2010) for the same image and filter and found to agree (Figure 2). The exact row of the image from which the blue filter data in Wolff et al. (2010) originated was not specified, beyond mention of it being near the location of Spirit Rover, so a row was chosen close to the position of Spirit at the time of image capture. The discrepancy in rows of data being compared means our calibrated data is not able to be an exact match to that of Wolff et al. (2010), but it is in close agreement. As in Figure 1 of Wolff et al. (2010), the blue filter data were multiplied by a factor of 0.75, and so was our data for this comparison.

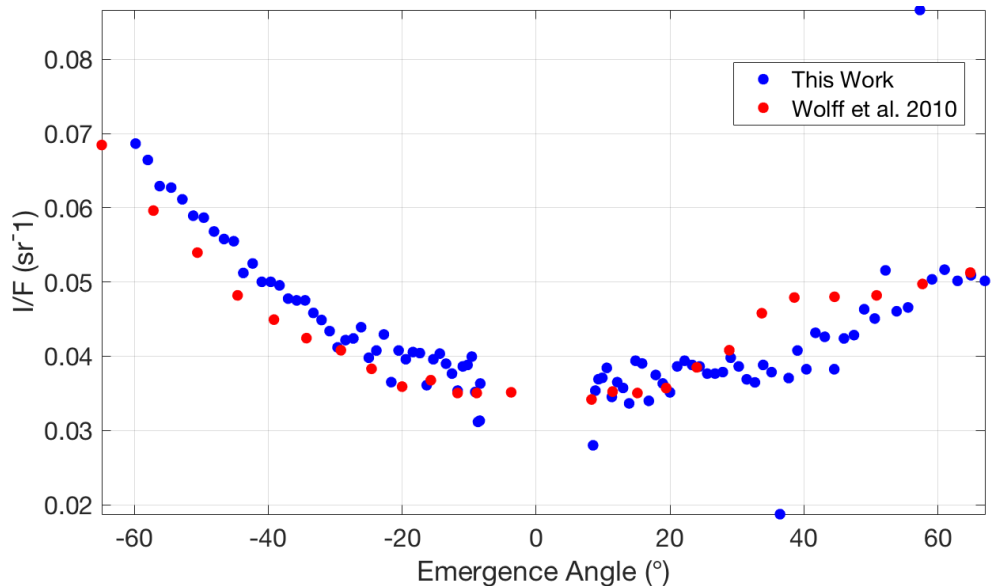

*Figure 2: The cross-track reflectances produced using our methods along a single row from the image "P10_004770_2885_MU_00N178W" were plotted alongside the reflectances from a nearby row in the same image from Wolff et al. (2010) to validate our calibration pipeline. The calibrated blue reflectances were multiplied by a factor of 0.75 in Figure 1 of Wolff et al. (2010), and the same was done to our data for a better comparison.*

**2.2 Phase Function Determination**

Equation 4 of Wang et al. (2014) approximates the scattering phase function, $P(\Theta, M)$, as a function of scattering angle, $\Theta$, and cloud microphysics, M, of a cirrus-type cloud observed from orbit with multiple single scattering:

$$P(\Theta, M) \approx \frac{4 R_{c\lambda} \times (\mu + \mu_0)}{\omega_o} \left[1 - \exp\left(-\frac{\tau(\mu + \mu_0)}{\mu \mu_0}\right)\right]^{-1} \quad (1)$$

where $\mu$ and $\mu_0$ are the cosine of the emission and Solar zenith angles of the cloud respectively (depicted in Figure 3), $\omega_o$ is the single scattering albedo which can be taken as 1.0 (Clancy et al., 2003), and $\tau$ is the cloud opacity. $R_{c\lambda}$ is the corrected cloud reflectance and accounts for two-way transmissivity, and bidirectional spectral reflectivity of the surface, $b_\lambda$, as per a re-arranged Equation 2 of Wang et al. (2014):

$$R_{c\lambda} = \frac{R_{obs\lambda} - b_\lambda - c_\lambda}{\exp\left(-\frac{\tau_a}{\mu}\right) \exp\left(\frac{\tau_a}{\mu_0}\right)} \quad (2)$$

where $R_{obs\lambda}$ is the calibrated reflectance observed from orbit, $c_\lambda$ is the spectral reflectivity of atmospheric gas, and $\tau_a$ is the opacity of the dust and ice hazes above the clouds. We used these two relations in combination to determine the phase function of the WICs in the MARCI image data.

Following the calibration detailed in Section 2.1, the maximum reflectance ($R_{obs\lambda}$ in the above notation) from each column of each single-filter image, assumed to be due to cloud regions, was recorded along with their respective indexed pixel location. To determine the central Martian longitude and latitude, and thus Solar incidence, emission, phase, and scattering angles, of these pixels-of-interest for use in Equation 1, a SPICE kernel algorithm was used.

SPICE is an information system produced by the Navigation and Ancillary Information Facility (NAIF), associated with NASA's Planetary Science Division (Acton, 1996). The information contained within the spacecraft, instrument, and ephemeris-based SPICE kernels was used in combination with multiple SPICE functions to determine and account for geometric distortion of the MARCI lens, and to provide geographic coordinates for the pixels of interest and their respective illumination angles.

The cloud opacities for use in Equation 1 came from integrated MCS water ice opacities (ranging from $0.050 \pm 0.001$ to $0.068 \pm 0.001$), binned for every 10° of $L_S$ (assuming uniform clouds over this range) and averaged over the range of latitudes and longitudes contained within the PSP data. They were converted from infrared to optical opacity by a conversion factor of 3.6 (Guzewich et al., 2017; Montabone et al., 2015; Kleinböhl et al., 2011). It should be noted that the MCS profiles have a mean profile lower limit of 13 km, and thus may not be capturing the complete cloud opacity. This is less of an issue with our focus on ACB clouds that typically extend from 10 km to 40 km in altitude (Clancy et al., 1996; Clancy et al., 2003). The ACB clouds were assumed to be capped at 40 km (Clancy et al., 1996; Clancy et al., 2003), and any ice opacity above was assumed to upper atmosphere ice haze.

The bi-directional spectral reflectivity of the surface, $b_\lambda$, came from the normalized 440nm phase function in Figure 27 of Johnson et al. (2006), spherically integrated and scaled to the mean spectral reflectance of the surface for the blue filter, from Figure 1 of Adams and McCord (1969). The reflectance from Adams and McCord (1969) was acquired through telescopic imagery taken near the center of the Martian disk to avoid any possible limb effects.

Only the blue filter data points that contained observed reflectances ($R_{obs\lambda}$) greater than their corresponding surface reflectivity were used in Equation 2. The spectral reflectivity of atmospheric gas was taken to be negligible for the MARCI visual filter band passes, thus the atmospheric gas above the clouds was assumed to be optically transparent. The opacity of the dust and ice hazes above the clouds for the two-way transmissivity correction were provided by the total summed ice and dust opacity (ranging from $0.001 \pm 0.001$ to $0.052 \pm 0.001$) from the seasonally averaged (+/- 5° $L_S$) MCS profiles above 40 km. The MCS infrared water ice opacity was converted to a visible opacity by multiplying by a factor of 3.6, and the dust opacity was converted using a factor of 7.9 (Guzewich et al., 2017; Montabone et al., 2015; Kleinböhl et al., 2011).

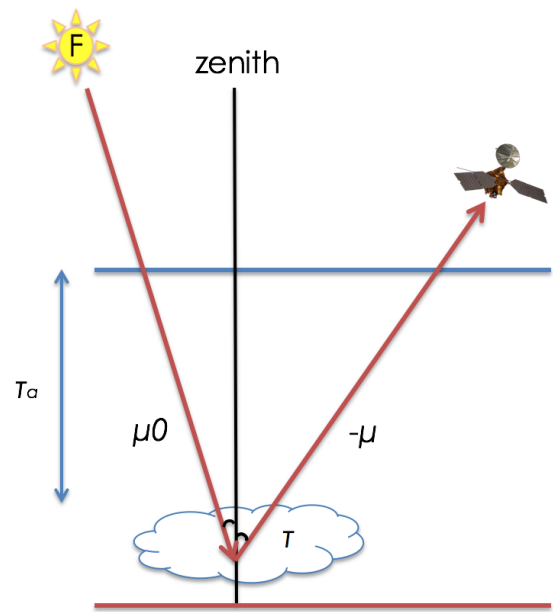

*Figure 3: A two dimensional diagram depicting the physical relevance of the cosine of the emission and solar zenith angle ($\mu$ and $\mu_0$) variables in Equation 1, the WIC opacity ($\tau$), and*

*the opacity of the dust and ice hazes above the clouds ($\tau_a$) in Equations 1 and 2. The sum of the emission and solar zenith angles is equal to the phase angle, and the scattering angle is equal to the phase angle subtracted from 180°.*

Even though the MARCI data in this analysis was constrained to the ACB seasons of MY 28 and 29, we cannot discount dust activity occurring during this time which would also contribute to the radiance observed from orbit. Battalio and Wang (2019) catalogued dust events stemming from the Aonia-Solis-Valles Marineris (ASV) region for MYs 24-31, which is a southern hemisphere dust storm track infrequently activated during $L_S$=120°-180° and is the source of the most impactful, organized dust activity during the ACB season. Once all the phase function values were produced from the MARCI data using the methods described above, they were compared to the times and locations of the dust events outlined in Battalio and Wang (2019) for MYs 28 and 29, filtering out the data points that overlapped temporally and spatially. The remaining phase function data points were used to produce average phase function curves for each VIS filter, and for seasonal and geographic phase function analyses in the blue and red filters.

The phase function is typically normalized to unity over all scattering angles (Greenler, 1980), so the shape of the phase function curve along various scattering angles becomes the relevant factor for comparison (Cooper et al. 2019) with other models (as opposed to the unnormalized absolute magnitude). The unnormalized phase function magnitudes are also useful, as they provide context for other analyses with respect to relative cloud opacity and wavelength-dependent reflectance. Standard phase function normalization can be difficult or impossible to do when the experimentally derived phase function cannot be determined over the entire range of scattering angles from 0° to 180°, as in this work. As such, we adopted the method of normalization from Cooper et al. (2019) that involved normalizing our derived mean phase function by the mean ACB phase function from Clancy et al. (2003) at the median scattering angle. The resultant normalized phase function curves were then used to constrain the seasonal dominant ice crystal geometries within the Martian WICs observed in this data set, and to probe the opposition surge.

## 3. Results

### 3.1 Blue Phase Function

Following the methods described in Section 2, the resultant phase function point density, average, and standard deviation for the blue filter over all solar longitudes, geographic latitudes, and longitudes are shown in Figure 4. The point density was produced by two-dimensionally binning the unnormalized phase function into 124 bins across both the observed scattering angle range (50.3° - 180.0°) and the unnormalized phase function magnitude. The number of points in each two-dimensional bin was then divided by the total number of points in its respective scattering angle column. The averages and upper bounds of the phase function data in each scattering angle bin were determined and smoothed using a simple boxcar averaging function with a width of three data bins, plotted overtop of the densities.

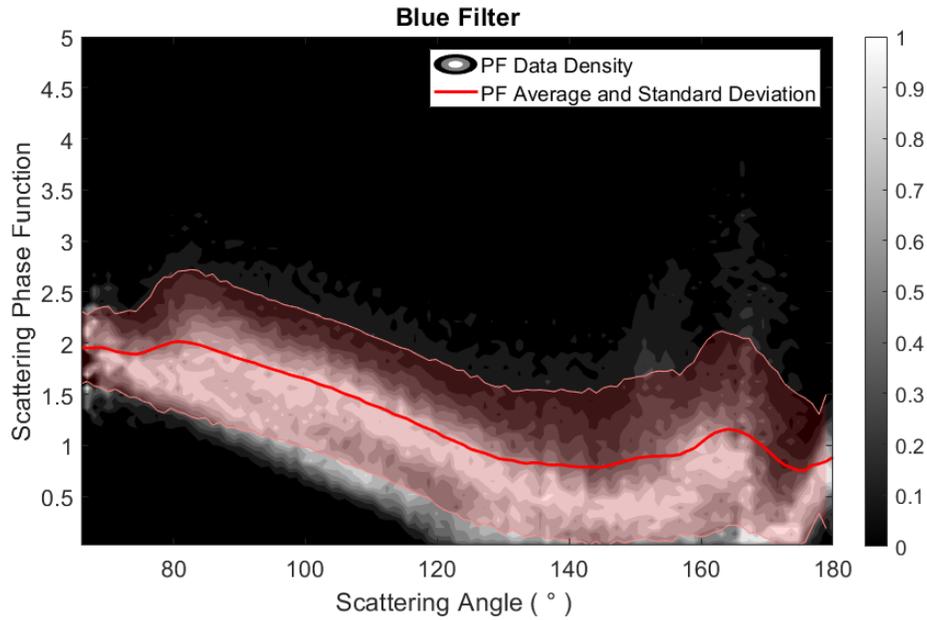

*Figure 4: The point densities of the filtered, unnormalized phase function data for the blue filter were produced by binning along both the scattering angle range and the phase function magnitude, and then dividing the number of data points in each two-dimensional bin by the total number of data points in its corresponding scattering angle column. The mean phase function curve and the one sigma variation of the point distributions along the scattering angle ranges were plotted overtop of the densities in red.*

As shown in Figure 4, the phase function curve has a convex upward shape, and the opposition surge is visible in the form of a peak leading up to the 180° scattering angle

### 3.2 Geographic Distribution

The filtered unnormalized phase function data points in the blue filter are shown in Figure 5 as a function of latitude and longitude (determined using SPICE for the centre of each 8x8 summed pixel region), in order to assess the geographic distribution and magnitude. The unnormalized phase function magnitude is dependent on both scattering angle and opacity, and thus panels (b-d) in Figure 5 contain data points within only a single scattering angle bin to isolate the magnitude effects of opacity (via calibrated pixel reflectance), which are eventually removed when the phase function is normalized in Section 4.3. As detailed in Section 2, the data points that were analyzed in our SPICE algorithm came from the radiometrically calibrated, summed, and cropped image pixels with maximum reflectance values in each column of each filter for each image. As a result, the location of the data points vary for each filter. From Figure 5, we see that the majority of the data points are acquired between -20° and +70° latitude, with two zonal bands around ~+20° and ~+40°. Fewer data points are acquired in the southern hemisphere compared the northern hemisphere, as MRO's orbit produces a seasonal offset in latitude coverage. This, in combination with uniform image cropping to exclude the polar caps at all times of year, led to data focused towards the northern hemisphere at the times of interest for this study. This offset is not of concern as it includes the primary latitudinal range of the ACB from -10° to +30° (Wolff et al., 1999; Clancy et al., 1996).

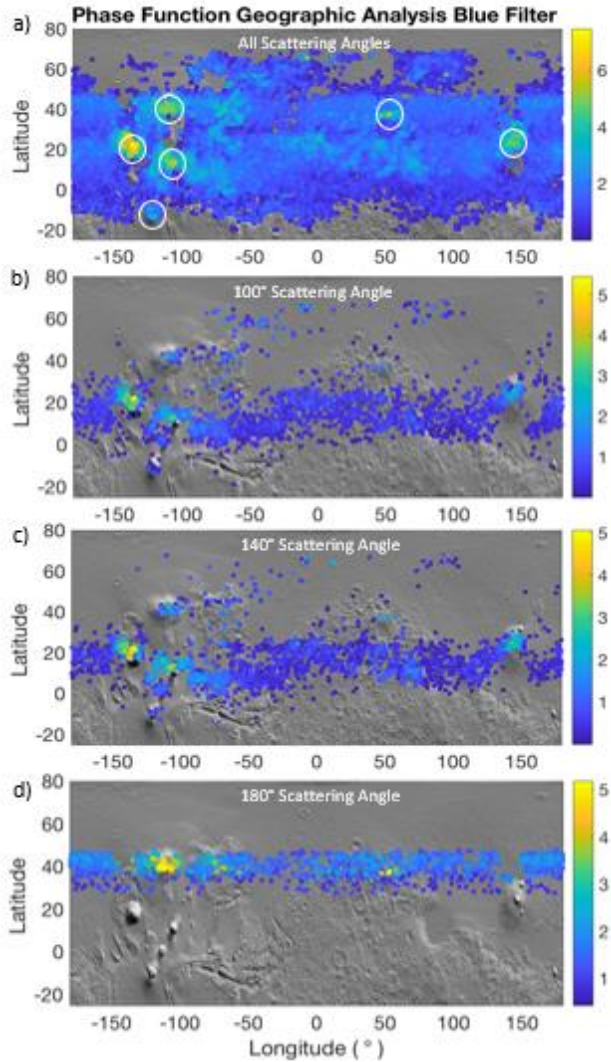

*Figure 5: The filtered and unnormalized phase function data points for the blue filter, shown with respect to their latitude and longitude. Panel (a) contains phase function data points from all scattering angles, while panels (b-d) contain phase function data points within the scattering angle bin listed. The color axis provides the unnormalized phase function magnitude, which provides some context regarding reflectance in panels (b-d). In both filters, where the peaks in magnitude (yellow points near Latitude 20°N and Longitude 140ºW) likely correspond to orographic clouds over Olympus Mons. Other peaks likely correspond to*

*orographic clouds around Alba Mons, Tharsis Montes, and Elysium Mons (circled in white in panel a).*

Looking at the distribution of phase function magnitudes in all panels of Figure 5, there are large magnitudes observed in around the locations of Olympus Mons, Tharsis Montes, Alba Mons, and Elysium Mons, indicating repeated detections of optically thick orographic clouds at these high elevation locations. The Mars Analysis Correction Data Assimilation (MACDA) (Montabone et al., 2014) shows the time-mean wind velocities at 2 pm at 15-40 km altitude between $L_S=141°-157°$ in MY 24 oriented in the south-eastward direction when examining the area to the south of Olympus Mons, Alba Mons and Elysium Mons (not shown). This suggests a leeward flow down the slopes of these mountains to their south, which could correspond to the lack of blue data points adjacent to these high magnitude points in panel (a), as the descending atmospheric parcels adiabatically warm and clouds sublimate. The zonal band centred on +20° latitude in Figure 5 (a) is located within the typical latitudes expected for the ACB (Wolff et al., 1999; Clancy et al., 1996), however the +40° latitude band extends further north, and corresponds to the single band in panel (d) caused by the opposition surge, likely from optically thin water ice hazes, as described in Cooper and Moores (2019). Figure 1 in Cooper and Moores (2019) depicts how each MARCI filter observes the 180° scattering angle (and thus the opposition surge) at a different point in orbit (and thus, different latitude), because each of the five VIS filters are permanently mounted on a different portion of the CCD. Panel (d) in Figure 5 shows only those points with a scattering angle of 180°, and thus includes only the points that correspond to the opposition surge. The opposition surge is discussed in more detail in Section 4.1.

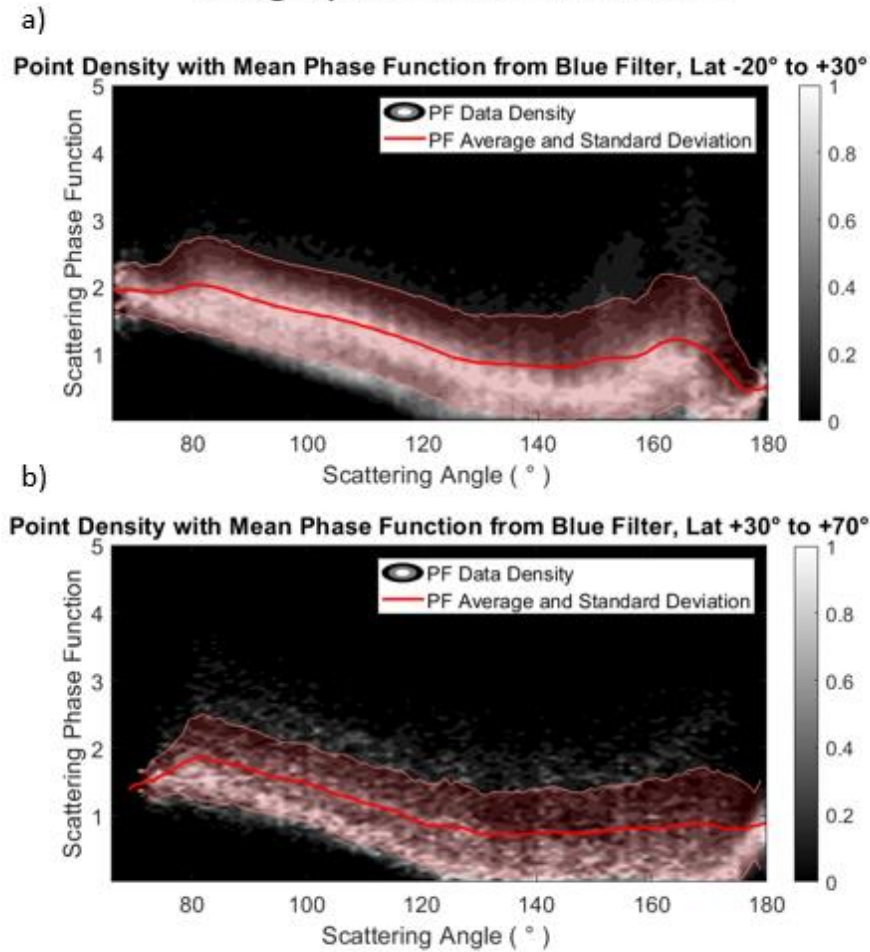

*Figure 6: The filtered unnormalized phase function point densities for the blue wavelength band were separated into two latitude ranges across all seasons: latitudes from -20° to +30° as the typical ACB range, and latitudes greater than +30° as the typical range of northern polar hood clouds and water ice hazes. Both the shape and distribution of points varies between the two latitude ranges. The 180° opposition surge is greater in data*

*corresponding to latitudes greater than +30°. The red line is the phase function mean, and the shaded range corresponds to one sigma variation.*

The unnormalized phase function data was divided into two latitude regions in Figure 6: data points at latitudes less than or equal to +30°, and data points at latitudes greater than +30°. Given that the majority of the data exists between -20° and +70° in latitude, the two ranges were chosen to separate the typical ACB latitudes (-10° to +30°) from those where northern polar hood (NPH) clouds (>+30°) and ice hazes are typically found.

The phase functions vary in both shape and width of standard deviation along all scattering angles, between the two latitude ranges. Referring back to panel (d) of Figure 5, it is clear that the 180° scattering angle is only observed by the blue portion of MARCI's CCD over a distinct latitude range, which extends from +27° to +46° over all seasons. Cooper and Moores (2019) showed that the observed scattering angles vary with latitude and filter, and furthermore change with seasonal illumination for a given latitude. As the range of latitudes in which the 180° scattering angle is observed extends into both the latitude ranges in Figure 6, we see the 180° peak in both. Panel (d) of Figure 5 shows the phase function values at a given scattering angle are higher for polar hood cloud latitudes than the ACB cloud latitudes. This agrees with the unnormalized phase function values in the 180° scattering angle bins of Figure 6, with a mean value of 0.96 for data at latitudes less than +30° and 1.6 for data at latitudes greater than +30°.

### 3.3 Seasonal Variation

The blue filter phase function in panel (a) of Figure 4 had data from all solar longitudes in the ACB seasons probed for this study, so in order to isolate the fluctuations caused by seasonal variations of cloud opacity, the investigation period was broken down into smaller solar longitude "seasons" for each MY. The blue filter was divided into solar longitude ranges of $L_S =$ 42°-84°, $L_S = 85°$-127°, and $L_S = 128°$-170° (because MRO reached Mars at ~$L_S = 128$ in MY28, only the last $L_S$ range could be investigated for that year). Phase function point densities were produced for each of the miniature seasons, along with an average curve and distribution upper-bound. The mean curves from each season were plotted overtop of their point densities in panels (a-d) Figure 7, and all together in panel (e) for a more direct comparison of their unnormalized magnitudes and shapes.

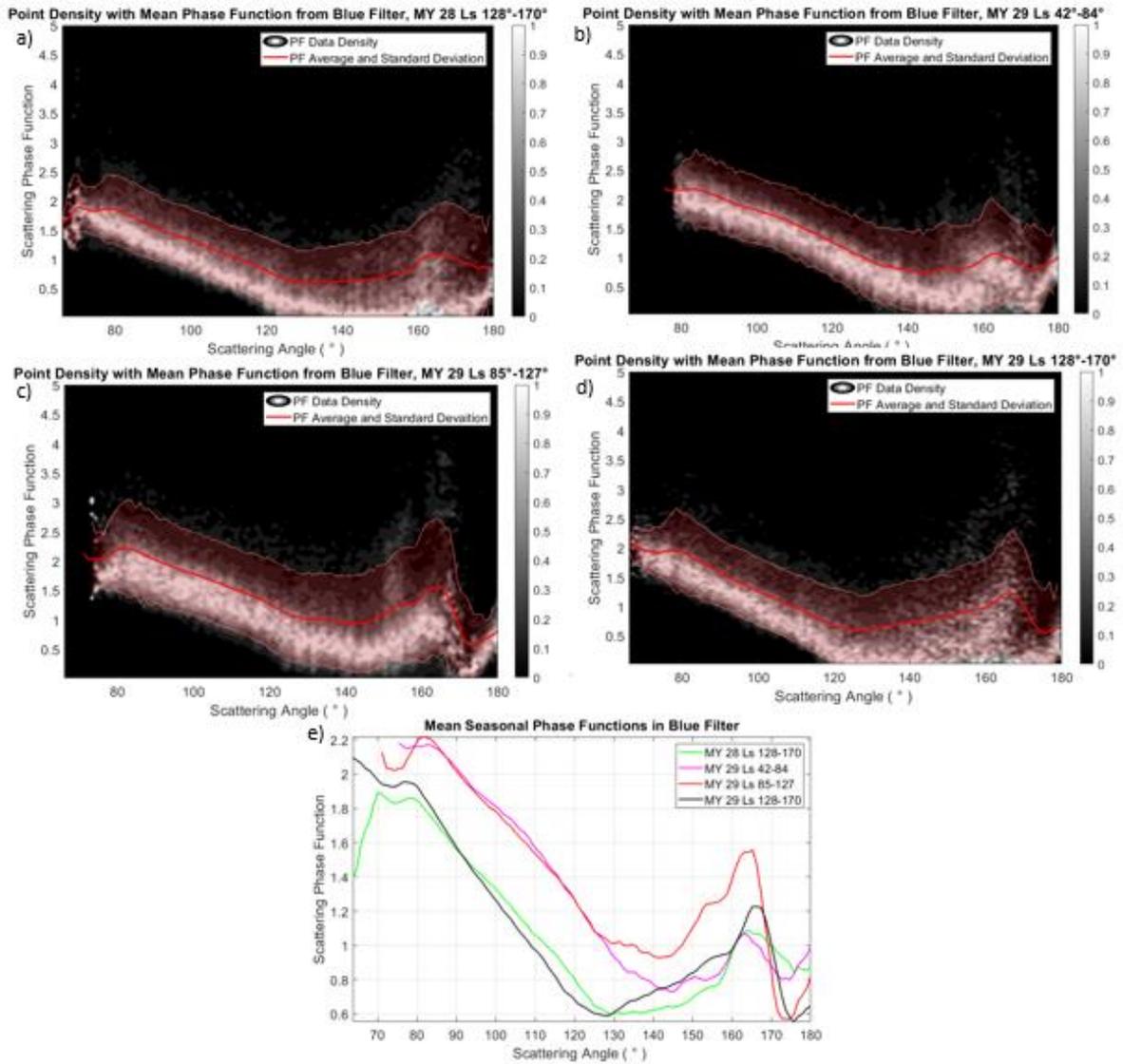

*Figure 7: The filtered unnormalized seasonal phase function point densities for the blue filter over solar longitude ranges of $L_S = 42°-84°$, $L_S =85°-127°$, and $L_S =128°-170°$ for MYs 28 and 29. The mean curves and standard deviation of the distributions are shown by the red and shaded regions in panels (a-d), and the mean curves shown together in panel (e).*

The MY 29 $L_S$=85°-127°, season had the most consistently high magnitudes with an average phase function magnitude of 1.0, followed by the MY 29 $L_S$ = 42°-84° season (0.92) and $L_S$=128°-170° from both MYs (0.80, and 0.82, respectively).

Figure 8 shows all filtered blue wavelength data across all seasons. In panel (c), data points at a single scattering angle of 165° are shown, and we can see that the majority of data points correspond to ACB clouds along all solar longitudes. This is the likely cause of the peak around the 165° scattering angle, and would explain why the peak is not present in the phase function for data at latitudes greater than +30° in Figure 6, where the 165° scattering angle corresponds to lower opacity WICs beyond the typical ACB range. It also aligns with the variation in magnitude of the ~165° peak in the seasonal phase functions of panel (e) Figure 7.

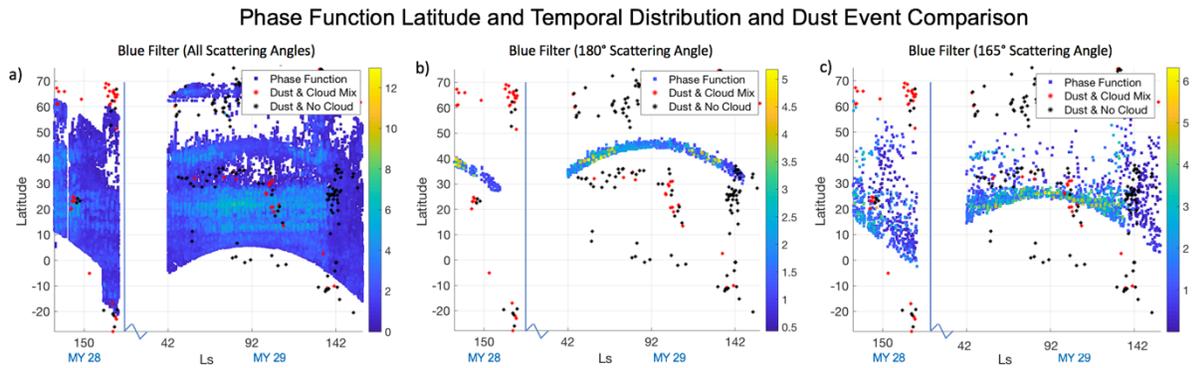

*Figure 8: The resultant unnormalized and unfiltered phase function values are plotted with respect to Martian latitude and solar longitude over the periods of interest in MYs 28 and 29. Panel (a) contains data within all the scattering angle bins, while the remaining panels contain data points within the scattering angle bin listed. The colour axes correspond to the phase function values for each filter, and the overlaid black and red points correspond to the*

*centres of dust events (without observed clouds, and with an observed dust-cloud mix, respectively) catalogued by Battalio and Wang (2019).*

## 4. Discussion

### 4.1 A Closer Look at the Opposition Surge

The opposition surge, caused by coherent backscatter, occurs at the 180° scattering angle that is frequently captured in MARCI images, and is amplified by the presence of Martian WICs to the point that it can produce an artifact with a rainbow-like appearance. This is captured when multiple MARCI VIS single filter images are combined to produce a false-colour composite RGB image (Cooper and Moores, 2019, and examples can be seen in Figure 9 of this work). Thus, the 180° peak should be represented in the phase function of Martian WICs as it is the only scattering phenomenon we have been able to observe from water ice crystals suspended in the Martian atmosphere.

The opposition surge was probed by a sensitivity study of the HWHM of the observed feature in the blue filter to minimize contributions from the surface, as in Cooper and Moores (2019). Five images (P22_009491_1083_MA_00N201W, P22_009496_1084_MA_00N337W, P22_009500_1086_MA_00N087W, P22_009514_1091_MA_00N109W, and P22_009522_1094_MA_00N327W, also shown in Figure 9) were randomly chosen from the middle of the ACB season, and calibrated using the methods outlined in Section 2.

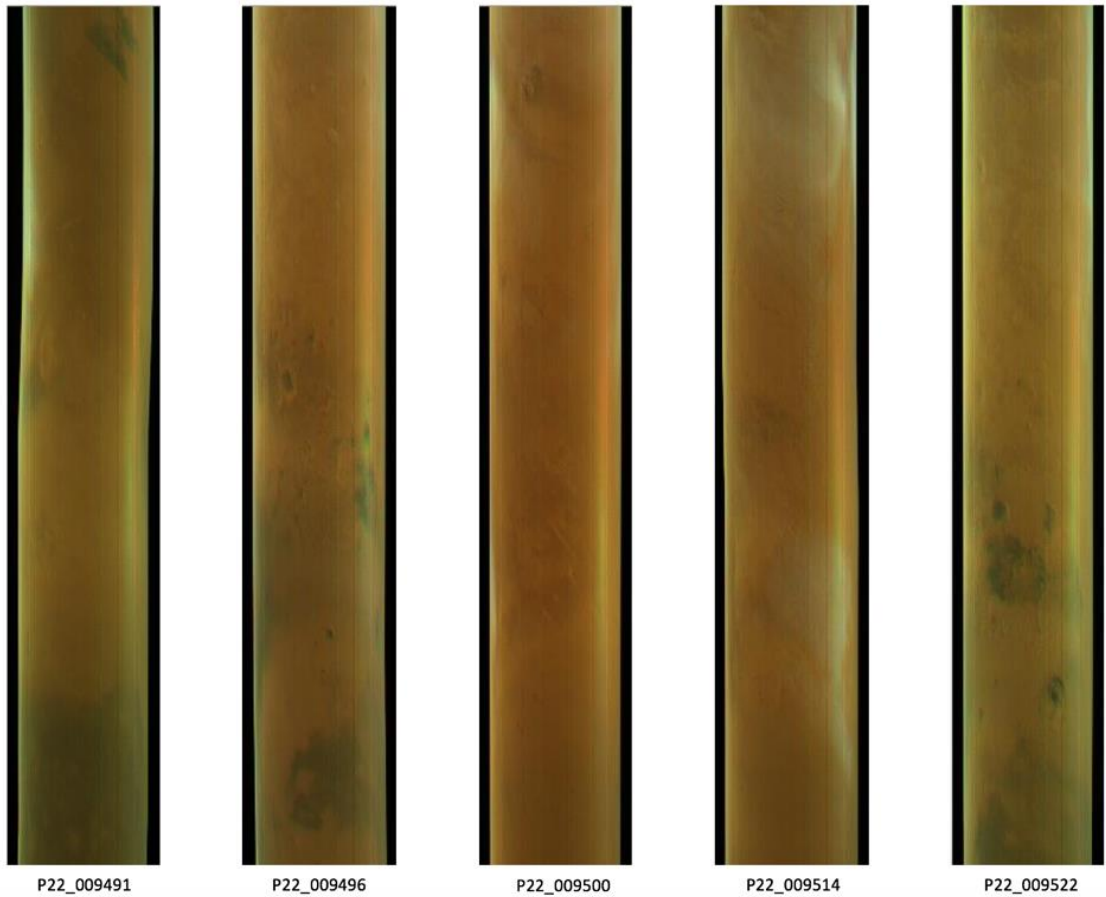

*Figure 9: Cropped RGB MARCI images from the ACB season are shown, with the commonly observed "rainbow" artifact that is produced by the opposition surge in different filters at different latitudes (Cooper and Moores, 2019). Images used: P22_009491_1083_MA_00N201W, P22_009496_1084_MA_00N337W, P22_009500_1086_MA_00N087W, P22_009514_1091_MA_00N109W, and P22_009522_1094_MA_00N327W.*

Reflectances and SPICE-derived latitudes along the opposition surge were used to measure the HWHM by fitting the peak in reflectance for each filter to a gaussian curve; this was done for all five images only in the blue filter to minimize contributions from the surface. The resultant HWHM were found to be (in the same order as the images were listed above): 3.94°, 1.95°, 3.14°, 1.66°, 1.93°. This led to an average HWHM of 2.52°, compared to the value of 3.42° from the single example in Cooper and Moores (2019). A great deal of cloud activity was visible in the Mars Global Daily Maps for all these images (MGDMs; Wang et al. 2018), which suggests a strong presence of ice aerosols are producing this overtly strong backscatter effect. This explanation is further justified by the fact that the HWHM of the opposition surge is within the range of values expected for WICs (approximately 2°- 5°; Chepfer et al., 2002; Yang and Liou, 1996; Yang et al., 2010), compared to values for the Martian dust or the surface (on the order of 10° or greater; Tomasko et al., 1999, Vincendon et al. (2014), and Soderblom et al., 2006).

**4.2 Does an Empirical Relationship Between Extinction and Ice Water Content Exist for the Martian Atmosphere?**

The Mars Phoenix lander was equipped with a lidar, and Dickinson et al. (2011) used the lidar derived water ice extinctions to estimate the ice water content (IWC) using an empirical relationship from in situ measurements of terrestrial cirrus clouds, that IWC (in units of $mg/m^3$) is equal to the WIC extinction (in units of $km^{-1}$) multiplied by a factor of 10. In order verify this empirical relationship, we used the mean phase function values in the 180° scattering bin for the blue filter (to minimize dust and surface influence), normalized in the method of Cooper et al. (2019). The mean normalized phase function value in the 180° scattering angle bin for the blue filter ($0.4 \pm 0.9\ sr^{-1}$) represents the observed backscatter from Martian WICs. This can be used in combination with the lidar ratio, the lidar equation, and Equation 1 from Moores et al. (2011), to test the relationship between IWC and integrated extinction (optical depth), in the atmospheric column. The resultant relationship between backscatter ($\beta$) and IWC is:

$$\beta \cong \frac{1}{4\pi}\left(\frac{3\ IWC}{2\ \alpha\rho}\right) \quad (3)$$

where $\alpha$ is the effective ice particle radius (a mean value of 2.75 microns for ACB clouds; Clancy et al., 2003), and $\rho$ is the density of water ice.

The lidar ratio ($S$) can be given by Equation 3 from Shin et al. (2018):

$$S = \frac{4\pi}{\omega_o\ F_{11}(180°)} \quad (4)$$

where $F_{11}(180°)$ is the element in the Müller scattering matrix at a scattering angle of 180°, by definition. Given the value of the single scattering albedo ($\omega_o$) of water ice clouds is already taken to be 1, and the mean value of our normalized derived 180° phase function is $0.4 \pm 0.9$ $sr^{-1}$, the resultant mean lidar ratio is 30 $sr$. The lidar ratio is also equal to the ratio of extinction ($\sigma$) and backscatter, thus the relationship between extinction ($\sigma$) and IWC can be re-written as:

$$\sigma \cong \frac{S}{4\pi}\left(\frac{3\ IWC}{2\ \alpha\rho}\right) \quad (5)$$

where all parameters are as defined previously.

Rearranging Equation 5 for IWC, we find that IWC is equal to the extinction multiplied by a mean factor of 5 for the mean 20 micron ice particles modeled to have been observed by Phoenix (Moores et al., 2011). The factor of 2 between our results and the relationship observed by Dickinson et al. (2011) was likely caused by the difference in the mean range of ice crystal sizes between the NPH and ACB, which is seen in the difference between our derived lidar ratios (also a factor of two). For the case of ACB ice crystals with a mean radius of 2.75 microns, the IWC was found to equal the extinction multiplied by a factor of 0.7. Given the range of magnitudes between these two factors, we can say that it is difficult to accurately derive the IWC from extinction for all ice crystals, as it requires knowledge of the mean radius of the ice crystals within the WICs observed, so stating a general empirical relationship for all sizes is not possible.

### 4.3 Crystal Habits of ACB Clouds

The five mean seasonal phase functions from Figure 7 (shown normalized in Figure 10 using the same normalization technique described earlier) were then compared to the seven modeled ice crystal phase functions from Yang and Liou (1996) and Yang et al. (2010), the observationally derived phase function from Cooper et al. (2019), and the six RT-fit phase functions from Clancy and Lee, (1991) and Clancy et al., (2003) within 0.3-2.7 micron and 0.3-3 micron bandpasses, respectively.

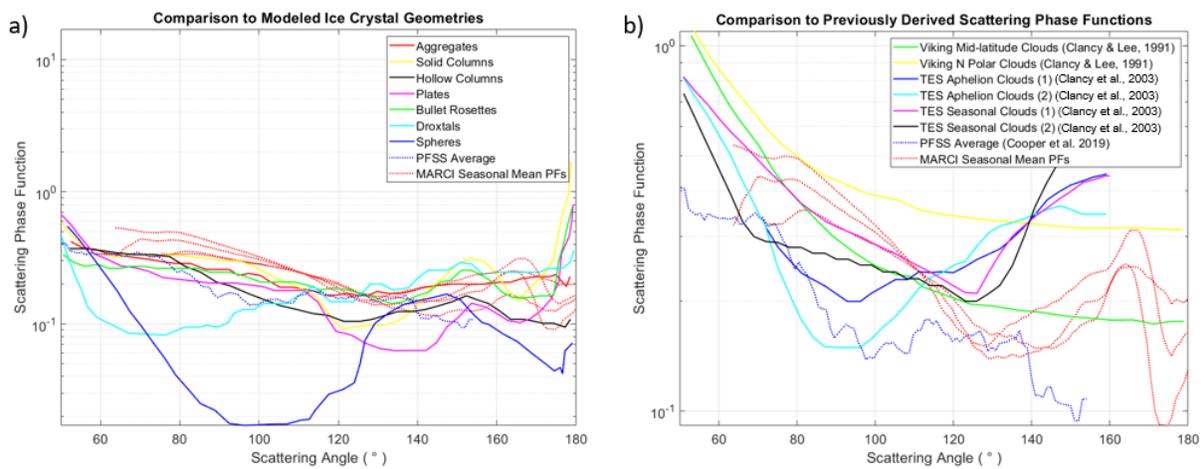

*Figure 10: The five seasonal mean phase functions from Figure 8 were normalized and are shown superimposed over the seven modeled ice crystal geometries with a maximum dimension of 50 microns from Yang and Liou (1996) and Yang et al. (2010) in panel (a). In panel (b), those same five phase functions are displayed with the observationally derived phase function from Cooper et al. (2019), and the six RT-fit phase functions from Clancy and Lee, (1991) and Clancy et al., (2003) to provide context (note the axis range differences between panels a and b).*

Figure 11 shows the results of the weighted chi-squared goodness of fit tests for the models from Yang and Liou (1996) and Yang et al. (2010), for the null hypothesis that there is no correlation between our derived phase functions and the modeled and previously derived phase functions. The chi-squared test was weighted by the degrees of freedom, which is equal to the number of columns minus one, multiplied by the number of rows minus 1. The number of rows was always two, as it corresponded to our derived phase function and the modelled phase function. The number of columns was equal to the number of scattering angle bins (out of the possible 124) that contained modeled/previously derived phase function values within them (and thus were able to be compared). This varied for some of the phase functions that were not derived over the entire investigation range (but was never greater than 124). For all seasons, the ice crystal geometries that were plausibly observed within the MARCI data were aggregates, bullet rosettes, plates, hollow and solid columns, and droxtals, while spheres were less plausible, but still possible.

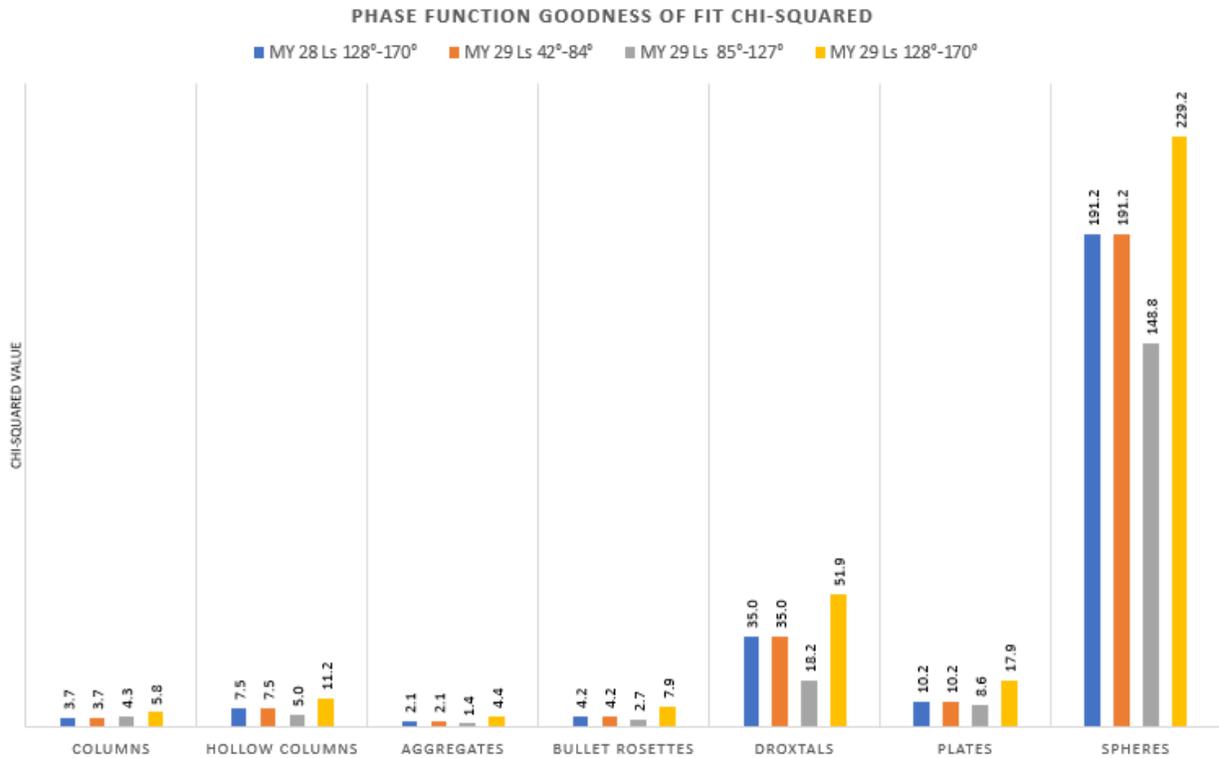

*Figure 11: The normalized mean seasonal phase functions were compared to seven modeled ice crystal phase functions from Yang and Liou (1996) and Yang et al. (2010) via weighted chi-squared analyses.*

From Whiteway et al. (2004), and Yang et al. (2003), in a typical Terrestrial WIC, ice crystal formation occurs at the top of the cloud, then the ice crystals grow, fall, and even sublimate as virga (observed in polar regions on Mars by Whiteway et al., 2009). Whiteway et al. (2004) sampled Terrestrial cirrus clouds over Australia and found a mix of rosettes at the highest levels along with columns and irregular shapes, while larger aggregates and irregular shaped crystals were present at the bottom of the clouds. Yang et al. (2003) discusses sampled mid-latitude cirrus and found small droxtals at the highest levels with temperatures below 223 K, pristine plates and columns in the mid-levels, and larger rosettes and aggregates at the lowest levels.

Both Whiteway et al. (2004) and Yang et al. (2003) suggest that we should see a range in ice crystal shapes and habits within Martian WICs, and given their low opacities, we should be able to observe all types from MARCI's orbital perspective. Our weighted chi-squared analysis corroborates that this is plausible.

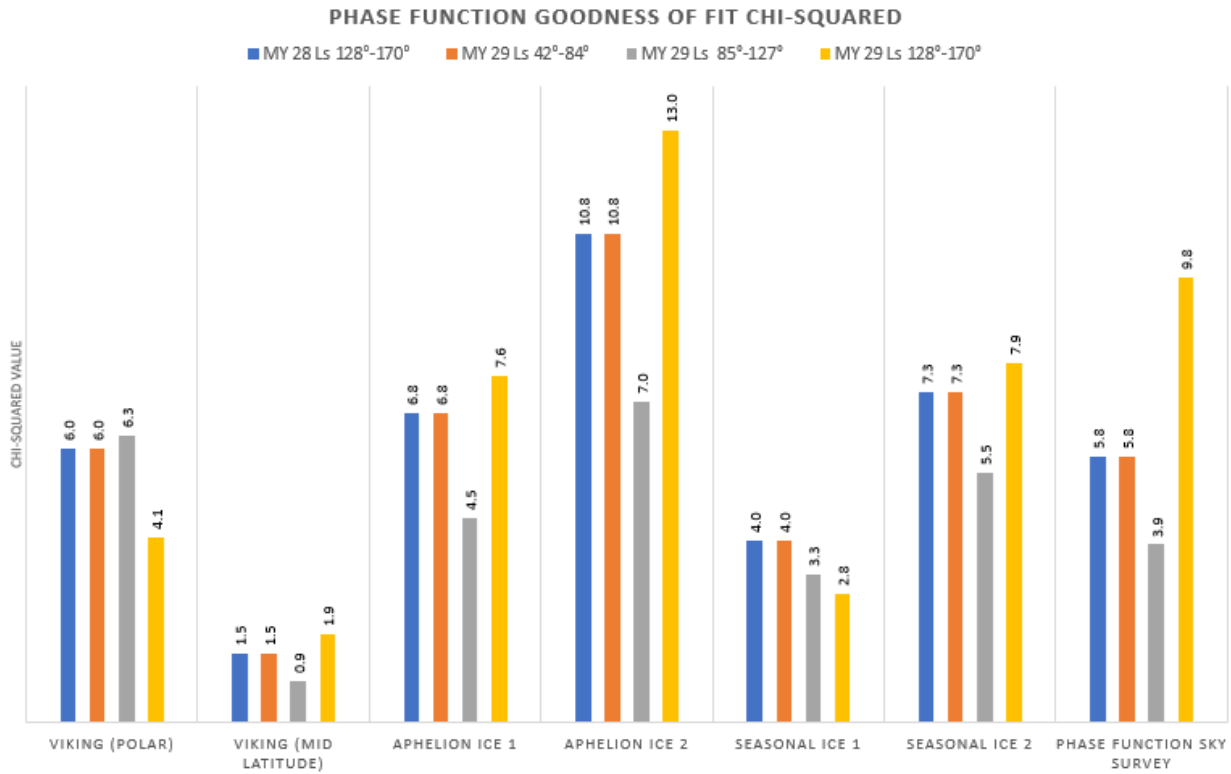

*Figure 12: The normalized seasonal phase functions were compared to the mid-latitude and polar phase functions from Viking EPFs (Clancy and Lee, 1991), aphelion and seasonal ice aerosol phase functions from Clancy et al. (2003), and the aphelion phase function sky survey results from Cooper et al. (2019) via weighted chi-squared analyses.*

The normalized mean seasonal phase functions were also compared to the mid-latitude and polar phase functions from Viking EPFs (Clancy and Lee, 1991), the aphelion and seasonal ice aerosol phase functions from Clancy et al. (2003), and the aphelion phase function sky survey (PFSS) results from Cooper et al. (2019). The results of the weighted chi-squared analysis are shown in Figure 12, showing that no habit can be ruled out, and that ice crystals predicted to produce phase functions within Clancy et al. (2003) and Clancy and Lee (1991) are plausibly contained within Martian WICs.

In Clancy et al. (2003), the aphelion (type 2) ice aerosol is postulated to be a "spheroidal" geometry, while the seasonal ice (type 1), is postulated to have a more "crystalline" shape. The droxtal habit would likely better represent the aphelion/type 2 ice from Clancy et al. (2003), as Yang et al. (2003) states that it is a more realistic crystal geometry for small ice crystals hypothesized to be "quasi-spherical", while still maintaining an aspect ratio of unity. Clancy et al. (2003) also puts forward the possibility of the type 1 aerosol taking the form of ice Ic, a variant from the typical form of ice Ih on Earth which produces the hexagonal-based shapes modelled by Yang and Liou (1996) and Yang et al. (2010). Ice Ic is a form of water ice where the oxygen atoms are arranged in a cubic or diamond structure and can lead to isometric forms like octahedrons or dodecahedrons. Gooding et al. (1986) investigated the role of dust particles in the atmosphere for deposition of water vapour to form WICs. They found that vapour/solid transitions in the Martian atmosphere favour ice Ic formation, and that the overall best substrate for the nucleation of other condensates is ice Ic, therefore, the most effective mineral substrates for condensate formation on Mars might be those that are most effective at nucleating ice Ic. It's additionally possible that we could be observing both water ice Ic and Ih, with one potentially nucleating the other.

Unfortunately, the viewing geometry and orbit of MRO during the PSP prevents us from observing the behaviour of MARCI-derived phase function curves at scattering angles below 60°. The shapes of the modeled curves at these smaller angles are much more varied and can help to better distinguish the prevalent geometries in the WICs observed. The fact that the MARCI phase function data was derived over a range of scattering angles where the majority of the phase functions converge makes it difficult to rule out any one particular geometry. The derived phase functions from this work require that the majority of the ice crystals within Martian WICs are irregular, or regular without resonance, thus precluding the formation and/or observation of halos, parhelia and other scattering phenomena.

## 5. Summary and Conclusions

The annual recurrence of the ACB within -10° and +30° latitude (Wolff et al., 1999; Clancy et al. 1996) contributes to the global radiation budget of Mars, and modelling the impact of these clouds from a radiative transfer perspective requires confidence in average cloud opacity, ice crystal habit, and particle size. While opacity is regularly monitored from the surface and orbit, constraining the shapes of the ice crystals is more difficult, but can be done by first constraining the scattering properties of the clouds via derivation of their average phase function. The goal of this work was to do just that by using publicly available data from the Mars Color Imager aboard MRO, during the primary science phase of the mission. Cooper et al. (2019) observationally constrained the phase function of Martian WICs from the surface of Gale crater using MSL, and this work extended that by allowing for the observation of clouds over a globally expanded range of Martian longitudes and latitudes over two MY ACBs. This work also probed the 180° peak (also known as the opposition surge) to validate the use of a Terrestrial empirical relationship for deriving IWC from water ice extinction or opacity from Dickinson et al. (2011), and to compare the 180° HWHM to those modeled by Chepfer et al. (2002), Yang and Liou, (1996) and Yang et al. (2010) for water ice crystals.

MARCI VIS filter data were calibrated and subsequently processed to select the pixels most likely to possess clouds and calculate their phase functions. The results were then compared to a dust event catalogue and analysis from Battalio and Wang (2019) to filter out any overlapping points and reduce contributions from dust in our data. The investigation period covered seasons $L_S$=42°-170° in MYs 28 and 29 (MRO reached Mars at ~$L_S$=128° in MY28), and unnormalized phase function data points were plotted with respect to scattering angle for MARCI's blue filter. The data was also utilized for the opposition surge analyses of Section 4.1 in order to compare to the blue MARCI filter 180° HWHM of Cooper and Moores (2019).

In testing the empirical relationship from Dickinson et al. (2011) that IWC is equal to a factor of 10 multiplied by the water ice extinction, we utilized the mean phase functions from the 180° scattering angle bins for the blue filter and found that the column IWC was equal to opacity multiplied by a factor of 5 for the average 20 micron particles at the Phoenix landing site, and 0.7 for the average 2.75 micron particles of ACB clouds. Thus, we found that there is a large dependence of IWC on particle size, and so a general empirical relationship for all ice crystals on Mars is not applicable. For the HWHM of the opposition surge investigations, we had results within the range of values expected for WICs (approximately 2°- 5°; Chepfer et al., 2002; Yang and Liou, 1996; Yang et al., 2010), compared to values for the Martian dust or the surface (on the order of 10° or greater; Tomasko et al., 1999, Vincendon et al., 2014, and Soderblom et al., 2006).

A seasonal ice crystal habit analysis was completed to determine how the phase function changed with solar longitude (averaged over all latitudes and longitudes) by dividing the data into seasons ($L_S$=42°-84°, $L_S$=85°-127°, $L_S$=128°-170° for each MY). The mean seasonal blue filter phase functions were normalized and compared to the seven modeled ice crystal phase functions from Yang and Liou (1996) and Yang et al. (2010), along with the observationally derived phase function from Cooper et al. (2019), and the six RT-fit phase functions from Clancy and Lee, (1991) and Clancy et al., (2003). None of the phase functions compared could be statistically rejected, and the derived phase functions required that the majority of the ice crystals within Martian WICs be irregular, or regular without resonance.

## 6. Acknowledgements

We would like to thank the MRO team for their efforts in making these data sets possible. Data was provided by the Navigation and Ancillary Information Facility and the Planetary Data System Imaging Node.

## 7. Funding

This work was supported by contributions from the Natural Sciences and Engineering Research Council (NSERC) of Canada's Collaborative Research and Training Experience Program (CREATE) for Technologies in Exo-Planetary Science (TEPS), as well as the Canadian Space Agency's Mars Science Laboratory Participating Scientist Program. J.M.B. is supported by NASA Mars Data Analysis Program (MDAP) grant 80NSSC17K0475.

## 8. Data Availability

Datasets (MRO-M-MARCI-2-EDR-L0-V1.0; Malin et al., 2001) related to this article can be found at https://pds-imaging.jpl.nasa.gov/data/mro/mars_reconnaissance_orbiter/marci/, on the PDS Imaging Node (Eliason et al., 1996).

Chepfer, Helene, Patrick Minnis, David Young, Louis Nguyen, and Robert F. Arduini. "Estimation of Cirrus Cloud Effective Ice Crystal Shapes Using Visible Reflectances from Dual-satellite Measurements." *Journal of Geophysical Research: Atmospheres* 107, no. D23 (2002). doi:10.1029/2000jd000240.

Clancy, R. Todd, Michael J Wolff., Philip R. Christensen "Mars Aerosol Studies with the MGS TES Emission Phase Function Observations: Optical Depths, Particle Sizes, and Ice Cloud Types versus Latitude and Solar Longitude." *Journal of Geophysical Research* 108, no. E9 (2003). doi:10.1029/2003je002058.

Clancy, R. Todd, and Steven W. Lee. "A New Look at Dust and Clouds in the Mars Atmosphere: Analysis of Emission-phase-function Sequences from Global Viking IRTM Observations." *Icarus* 93, no. 1 (1991): 135-58. doi:10.1016/0019-1035(91)90169-t.

Clancy, R.T., A.W. Grossman, M.I. Wolff, P.B. James, D.I. Rudy, Y.N. Billawala, B.I. Sandor, S.W. Lee, and D.O. Muhleman. "Water Vapor Saturation at Low Altitudes around Mars Aphelion: A Key to Mars Climate?" *Icarus* 122, no. 1 (1996): 36–62. https://doi.org/10.1006/icar.1996.0108.

Cooper, Brittney, and John Moores. "A Surprising and Colorful Martian Scattering Artifact." Research Notes of the AAS, vol. 3, no. 2, 2019, p. 40., doi:10.3847/2515-5172/ab082a.

Cooper, Brittney A., John E. Moores, Douglas J. Ellison, Jacob L. Kloos, Christina L. Smith, Scott D. Guzewich, and Charissa L. Campbell. "Constraints on Mars Aphelion Cloud Belt Phase Function and Ice Crystal Geometries." Planetary and Space Science 168 (2019): 62-72. doi:10.1016/j.pss.2019.01.005.
37

Moores, John., Smith, P., Tanner, R., Schuerger, A., & Venkateswaran, K. (2007). The shielding effect of small-scale Martian surface geometry on ultraviolet flux. Icarus, 192(2), 417-433. doi:10.1016/j.icarus.2007.07.003

Newman, Claire E., Stephen R. Lewis, and Peter L. Read. "The Atmospheric Circulation and Dust Activity in Different Orbital Epochs on Mars." *Icarus* 174, no. 1 (2005): 135–60. https://doi.org/10.1016/j.icarus.2004.10.023.

Petrosyan, A., B. Galperin, S. E. Larsen, S. R. Lewis, A. Määttänen, P. L. Read, N. Renno, L. P. H. T. Rogberg, H. Savijärvi, T. Siili, A. Spiga, A. Toigo, and L. Vázquez. "The Martian Atmospheric Boundary Layer." Reviews of Geophysics 49, no. 3 (2011). doi:10.1029/2010rg000351.

Poetzsch-Heffter, C., Q. Liu, E. Ruperecht, and C. Simmer. "Effect of Cloud Types on the Earth Radiation Budget Calculated with the ISCCP Cl Dataset: Methodology and Initial Results." Journal of Climate 8, no. 4 (1995): 829-43. doi:10.1175/1520-0442(1995)0082.0.co;2.

Pollack, James B., David S. Colburn, F. Michael Flasar, Ralph Kahn, C. E. Carlston, and D. Pidek. "Properties and Effects of Dust Particles Suspended in the Martian Atmosphere." *Journal of Geophysical Research* 84, no. B6 (1979): 2929. doi:10.1029/jb084ib06p02929.

Ruff, Steven W., & Christensen, P. R. (2002). Bright and dark regions on Mars: Particle size and mineralogical characteristics based on Thermal Emission Spectrometer data. Journal of Geophysical Research: Planets, 107(E12). doi:10.1029/2001je001580

Schlimme, I., A. Macke, and J. Reichardt. "The Impact of Ice Crystal Shapes, Size Distributions, and Spatial Structures of Cirrus Clouds on Solar Radiative Fluxes." Journal of the Atmospheric Sciences 62, no. 7 (2005): 2274-283. doi:10.1175/jas3459.1.41

Vincendon, M., Audouard, J., Altieri, F., & Ody, A. (2015). Mars Express measurements of surface albedo changes over 2004–2010. Icarus, 251, 145-163. doi:10.1016/j.icarus.2014.10.029

Wang, Chenxi, Ping Yang, Andrew Dessler, Bryan A. Baum, and Yongxiang Hu. "Estimation of the Cirrus Cloud Scattering Phase Function from Satellite Observations." Journal of Quantitative Spectroscopy and Radiative Transfer 138 (2014): 36-49. doi:10.1016/j.jqsrt.2014.02.001.

Wang, Huiqun, Michael Battalio, and Zachary Huber. MARS MRO MARCI Mars Daily Global Maps Archive | USGS Astrogeology Science Center. Accessed May 15, 2019. https://astrogeology.usgs.gov/search/map/Mars/MarsReconnaissanceOrbiter/MARCI/MARS-MRO-MARCI-Mars-Daily-Global-Maps.

Wang, Huiqun, & Richardson, M. I. (2015). The origin, evolution, and trajectory of large dust storms on Mars during Mars years 24–30 (1999–2011). Icarus, 251, 112-127. doi:10.1016/j.icarus.2013.10.033

Whiteway, J. A. and 23 Co-Authors. "Mars Water-Ice Clouds and Precipitation." *Science* (2009) 68–70. doi: 10.1126/science.1172344.

Whiteway, James, Clive Cook, Martin Gallagher, Tom Choularton, John Harries, Paul Connolly, Reinhold Busen, Keith Bower, Michael Flynn, Peter May, Robin Aspey, and Jorg Hacker. "Anatomy of Cirrus Clouds: Results from the Emerald Airborne Campaigns." *Geophysical Research Letters*, vol. 31, no. 24, 2004, doi:10.1029/2004gl021201.

Wolff, Michael J., R. Todd Clancy, Jay D. Goguen, Michael C. Malin, and Bruce A. Cantor. "Ultraviolet Dust Aerosol Properties as Observed by MARCI." Icarus 208, no. 1 (2010): 143-55. doi:10.1016/j.icarus.2010.01.010.
43